\documentclass[prd,aps,showpacs,eqsecnum,floatfix,nofootinbib,preprint,tightenlines]{revtex4}

\usepackage{latexsym}
\usepackage{epsf}
\usepackage{graphicx}
\usepackage{slashed}
\usepackage{amsmath}
\usepackage{amssymb}

\def\g{\gamma}

\def\p{\pi}                      

\def\G{\Gamma}

\def\Tr{{\rm Tr}\,}

\def\cbo{{\,\raise-.15ex\Sc [\,}}                       

\def\ddt#1{{\buildrel {\hbox{\LARGE .\kern-2pt.}} \over {#1}}}

\long\def\symbolfootnote[#1]#2{\begingroup%
\def\thefootnote{\fnsymbol{footnote}}\footnote[#1]{#2}\endgroup}

\long \def \blockcomment #1\endcomment{}

\def\Tr{{\rm Tr}}

\def\tq{{\tilde{q}}}

\def\vp{{\vec p}}

\def\vn{{\vec n}}

\newcommand{\pref}[1]{(\ref{#1})}

\newcommand{\mpi}{M_{\pi}}
\newcommand{\mN}{M_N}
\newcommand{\Nbar}{\overline{N}}
\newcommand{\Nbarpm}{\overline{N}_{\!\pm}}

\begin{document}
\hyphenation{fer-mio-nic per-tur-ba-tive pa-ra-me-tri-za-tion
pa-ra-me-tri-zed a-nom-al-ous}

\renewcommand{\thefootnote}{$*$}

\preprint{HU-EP-15/11}

\title{Nucleon-pion-state contribution to 
nucleon two-point correlation functions}

\author{Oliver B\"ar$^{a}$} 
\affiliation{$^a$Institut f\"ur Physik,
\\Humboldt Universit\"at zu Berlin,
\\12489 Berlin, Germany}

\begin{abstract}
We study the nucleon-pion-state contribution to the QCD two-point function of standard nucleon interpolating fields. For sufficiently small quark masses  these two-particle states are expected to have a smaller total energy than the single-particle excited states. 
We calculate the nucleon-pion-state contribution to leading order in chiral perturbation theory. Both parity channels are considered. We find the nucleon-pion-state contribution to be small, contributing at the few percent level to the effective mass in the positive parity channel.
\end{abstract}

\maketitle

\renewcommand{\thefootnote}{\arabic{footnote}} \setcounter{footnote}{0}

\newpage
\section{\label{Intro} Introduction}

Lattice QCD has made enormous progress over the last years due to computational and algorithmic advances \cite{Schaefer:2012tq}.
This has led to significantly improved lattice calculations of many low-energy observables. 
Present-day unquenched lattice calculations are performed with quark masses close to or at their physical value \cite{Abdel-Rehim:2014nka,Bazavov:2014wgs,Durr:2013goa,Aoki:2009ix}. Uncertainties associated with the chiral extrapolation are essentially eliminated in these simulations. 

However, dynamical lattice QCD with small quark masses may face new problems. One feature of unquenched lattice simulations is the presence of multi-particle states in the correlation functions measured to obtain observables. With the up and down quark masses getting closer to their physical values one expects multi-particle states with additional pions to become a significant excited-state contamination in many correlation functions.  As a simple example consider the two-point function $C(t)$ of a nucleon interpolating field, as it is measured to extract the nucleon mass $M_N$. 
From the spectral decomposition the two-point function in a finite spatial volume, projected to zero momentum, is a sum of exponentials,
\begin{equation}\label{ExpAnsatz}
C(t) = b_0 e^{-M_0 t} + b_1 e^{-M_1 t} + \ldots\,.
\end{equation}
The first exponential provides the exponential decay with the nucleon mass, $M_0=M_N$. All the other terms stem from states with the same quantum numbers as the nucleon, either genuine single-particle excited states or multi-particle states. For sufficiently small pion masses  one expects a nucleon-pion state to be the state with lowest total energy next to the ground state, $M_1\approx E_N + E_{\pi}$. For symmetry reasons the nucleon and the pion cannot be at rest. Both have non-zero but opposite spatial momenta  determined by the spatial volume and the boundary conditions imposed in the spatial directions. Still, for sufficiently large spatial volumes the exponent $M_1$ can be smaller than the first  one-particle excited state, associated with the Roper resonance $N^*(1440)$ in infinite volume. Moreover, near physical quark masses the three-particle state containing the nucleon and two pions at rest will have a smaller energy than the one-particle excited state.

The way to deal with multi-particle states in spectroscopy calculations is well-known. The well-established variational method \cite{Luscher:1990ck} can be used provided interpolating fields for the multi-particle states are taken into account.\footnote{Very recent studies that include two-particle nucleon-pion states in the analysis are reported in Refs.\ \cite{Verduci:2014csa,Kamleh:2014nxa,Kiratidis:2015vpa}, for example.} 
Still, the more states one takes into account  the larger is the generalized eigenvalue problem one has to solve numerically, and the error bounds for the energies derived in \cite{Blossier:2009kd} get worse the denser the spectrum is.

In this paper we provide some analytical results  concerning the nucleon two-point function. As has been pointed out in  Ref.\ \cite{Bar:2012ce}, chiral perturbation theory (ChPT) can be employed to obtain an estimate for the ratio $b_1/b_0$.\footnote{To our knowledge the idea for using ChPT to study the two-particle-state contributions to nucleon correlation functions was put forward first in Ref.\ \cite{Tiburzi:2009zp}.} 
Moreover, to leading order in the chiral expansion one expects the ratio $b_1/b_0$ to be independent of the a priori unknown low-energy constant (LEC) associated with the particular choice for the interpolating field. In that sense LO ChPT makes a rather definite prediction for $b_1/b_0$.  
Even if this result will receive substantial higher order corrections we do obtain a reliable first estimate for the impact of the nucleon-pion-state contribution to the two-point function.

The results we find for $b_1/b_0$ are small. For example, for a pion mass satisfying $\mpi L \approx 4$ and $\mpi/\mN\approx 0.2$ we find $b_1/b_0\approx 0.1$.  Taking into account the additional exponential suppression the two-particle-state contribution in \pref{ExpAnsatz} contributes at the few-percent level for euclidean times of about 0.5 fm. For larger $t$ and in the effective mass the contribution is even smaller. Whether it is noticeable in practice is then a question of the size of the statistical errors in the lattice data.

The nucleon-pion-state contribution to the two-point function has been independently computed in Ref.\ \cite{Tiburzi:2015tta}. The computation in that reference is performed in heavy baryon chiral perturbation theory (HBChPT), while here we employ the covariant formulation. In particular the chiral expressions for the interpolating fields differ in these two formulations, thus the final results are not the same. However, performing the appropriate expansion of our results we reproduce  the results given in Ref.\ \cite{Tiburzi:2015tta}. 

\section{\label{secQCD} Nucleon two-point correlators in QCD}

\subsection{General considerations}\label{ssect:general}

Throughout this article we consider QCD in a finite spatial box.  $L$ denotes the box length in  each direction and periodic boundary conditions are assumed. The euclidean time extent, however, is taken infinite. This choice implies an exponential decay of two-point functions and  simplifies our calculations. Still, this simplification is a good approximation for many lattice QCD simulations. Another simplification concerns the masses of the up and down type quarks which we assume to be equal. Consequently, all three pions as well as the nucleons (proton and neutron) are mass degenerate.

We are interested in the two-point correlation functions of a nucleon interpolating field $N$ with definite parity, 
\begin{eqnarray}
C_{\pm}(t)& =& \int_{L^3} {\rm d}^3{{x}}\, \langle  N_{\pm}(\vec{x},t) \Nbarpm(0,0)\rangle\ .\label{DefNCorr}
\end{eqnarray}
Here we defined $N_{\pm} = \G_{\pm} N$ and $\Nbarpm = \Nbar \G_{\pm}$ with the standard projectors 
$\G_{\pm} = (\g_0\pm 1)/2$.\footnote{Note that in order to keep the notation simple we often suppress the Dirac and/or flavor indices. For example, eq.\ \pref{DefNCorr} contains an implicit summation over the Dirac indices.} $N$ itself is an interpolating field with the quantum numbers of the nucleon. Various choices are possible and we discuss concrete examples in the next subsection. For the moment we do not need to specify $N$ any further. 

Let us consider the positive parity correlator.
The integration over the spatial volume in Eq.~\pref{DefNCorr} projects on states with zero total momentum. Hence, in the spectral decomposition of the correlator for large euclidean times $t\gg0$ the dominant contribution comes from the single-particle state with the particle being the nucleon at rest, 
\begin{equation}
\label{spcontr}
C_{+, N}(t)=
\frac{1}{2M_{\pm}}\;|\langle 0|N_{+}(0)|N(\vp=0)\rangle|^2 e^{-M_{N} t } \,.
\end{equation}
Here $|N(\vp=0)\rangle$ is the nucleon state 
and $\mN$ denotes the nucleon mass.

The interpolating field excites other states with the same quantum numbers as well. The contribution of an excited nucleon has the same form as Eq.~(\ref{spcontr}) with the appropriate mass $M'> M_{N}$. In addition we expect contributions from multi-hadron states. For sufficiently small pion masses the dominant multi-hadron states are those containing additional pions. For the two-particle nucleon-pion state contribution one finds
\begin{eqnarray}
\label{tpcontr}
C_{+, N\pi}(t)&=&\frac{1}{L^3}\;\sum_{\vp}\frac{1}{4E_{N} E_{\pi}}\,
|\langle 0|N_{+}(0)|N(\vp) \pi(-\vp)\rangle|^2 e^{-E_{\rm tot}t}\,.
\end{eqnarray}
$E_{\rm tot}$ is the total energy of the state and $E_N$, $E_{\pi}$ are the individual energies of the nucleon and the pion, respectively. For weakly interacting pions $E_{\rm tot}$ equals approximately the sum $E_N+E_{\pi}$. The sum over momenta runs over all momenta allowed by the boundary conditions imposed for the finite spatial volume, e.g.\ $\vp=2\p\vn/L$ with $\vn$ having integer-valued components. 

A simple dimensional analysis can be employed to make the volume suppression more quantitative. 
Assume the interpolating fields are local 3-quark-operators without derivatives.
In that case the mass dimension of the matrix elements in \pref{spcontr} and \pref{tpcontr} are 7/2 and 5/2, respectively. Making the naive assumption $\langle 0|N_{\pm}(0)|N(\vp) \pi(-\vp)\rangle\approx \langle 0|N_{\pm}(0)|N_{\pm}(\vp=0)\rangle / f_{\pi}$ we can estimate the ratio of the two-particle and one-particle contributions as
\begin{equation}\label{NaiveEst}
\frac{C_{+, N\pi}(t)}{C_{+,N}(t)}\approx \frac{1}{2(f_{\pi}L)^2M_{\pi}L}\frac{M_{\pi}}{E_{\pi}}\frac{M_{N}}{E_{N}}e^{-(E_{\rm tot} - M_{N})t}\,.
\end{equation}
If we assume the values $M_{\pi}\approx 200$ MeV and $L\approx 4$ fm we roughly find $[2(f_{\pi} L)^2M_{\pi}L]^{-1}\approx 1/30$. The additional factors suppress the two-particle state contribution further, so we expect its contribution to the correlator to be rather small. 

States with more than one pion contribute analogously to \pref{tpcontr}, but each additional pion contributes an additional factor $[2(f_{\pi} L)^2M_{\pi}L]^{-1}$, i.e.\ the more pions in the state the larger the suppression of its contribution with the spatial volume.

Our discussion applies to the negative parity correlator as well. In this channel the lightest single-particle state is, in infinite volume,  the $N^*(1535)$. However, the state with the lowest energy is the nucleon-pion state with both particles at rest, provided the pion mass is sufficiently small. Thus the two-particle state dominates the long time behaviour and the ratio analogous to the one in \pref{NaiveEst} will diverge for $t\rightarrow \infty$ in the negative parity channel. 

\subsection{Interpolating fields for the nucleon}\label{ssect:QCDinterpolaters}

As already mentioned, there exist many choices for the interpolating field (``operator'') $N$ with the quantum numbers of the nucleon. 
The number is significantly reduced if we consider local operators composed of three quark fields at the same point $x$. If, in addition, we constrain ourselves to operators  without derivatives there exist only five different ones. As a consequence of Fierz identities only two are independent \cite{Ioffe:1981kw,Espriu:1983hu} (see also \cite{Nagata:2008zzc}) and we focus on those. In order to write them down it is convenient to introduce the quark field doublet $\tq$ as 
\begin{equation}\label{qtilde}
\tq=q^{\rm T} C \gamma_5 (i\sigma_2)\,.
\end{equation}
Here $q= (u,d)^{\rm T}$ is the isospin doublet of the quark fields, $C$ denotes the Dirac spinor charge conjugation matrix satisfying $\gamma_{\mu}^{\rm T} = -C\gamma_{\mu} C^{-1}$, and $\sigma_2$ is the second (isospin) Pauli matrix. With these definitions the two nucleon operators can be written as
\begin{equation}\label{DefN12}
\begin{array}{l}
 N_1 \,= \, (\tq q) q\,,\\[0.4ex]
 N_2\,=\, (\tq \gamma_5 q) \gamma_5q\,.
 \end{array}
\end{equation}
This compact form suppresses the contraction of the isospin and Dirac indices in the bilinear quark fields $(\tq q)$ and $(\tq \gamma_5 q)$ (``diquarks'') and the summation over the color indices with an $\epsilon_{abc}$ to form a color singlet. The nucleon operators $N_i$ are still isospin doublets. To project onto the quark content of the proton and neutron we need to contract with the isospin basis vectors $e_p=(1,0)^{\rm T}$ and $e_n=(0,1)^{\rm T}$, respectively. However, in our case with preserved isospin symmetry any unit vector would be equally good.

In the next section we need the counterparts of the nucleon operators $N_i$ in ChPT. The mapping follows the standard procedure and rests on the transformation properties of $N_i$ under chiral and parity transformations. 
The transformation properties under (singlet and non-singlet) chiral transformations have been studied in detail in Ref.\ \cite{Nagata:2008zzc}. Here we simply summarize the relevant results. 

We decompose the quark fields into right- and left-handed components, $q=q_R+q_L$, with the usual chiral projectors $P_{+}=(1+\gamma_5)/2$ and $P_{-}=(1-\gamma_5)/2$. It then follows that the field in \pref{qtilde} also decomposes according to $\tq=\tq_R+\tq_L$, with $\tq_R=\tq P_+$ and $\tq_L=\tq P_{-}$. The group of non-singlet chiral transformations is $G=SU(2)_R\otimes SU(2)_L$, and under transformations $R\otimes L\in G$ the chiral quark fields transform according to
\begin{equation}\label{chiralTrafoq}
\begin{array}{rclcl}
q & =&  q_R + q_L &  \xrightarrow{\,R,L\,} & R \,q_R + L  \,q_L\,,\\[0.4ex]
\tilde{q} & =&  \tilde{q}_R + \tilde{q}_L & \xrightarrow{\,R,L\,} & \tilde{q}_R \,R^{\dagger}  + \tilde{q}_L \,L^{\dagger}\,.
\end{array}
\end{equation}
The diquarks decompose into $\tq q =\tq_R q_R + \tq_Lq_L$  and $\tq \gamma_5 q=\tq_R q_R - \tq_Lq_L$, hence they transform as singlets under chiral transformations. Consequently, the transformation behavior of the nucleon operators is determined by the third quark field contribution $q$ and $\gamma_5 q$, given in \pref{chiralTrafoq}. Decomposing this quark field into right- and left-handed components the complete nucleon fields $N_{1,2}$ itself can be written as a sum of a right-handed and a left-handed term with the following transformation behavior under chiral transformations:
\begin{equation}\label{NTrafoLR}
\begin{array}{rclcl}
N_{i} & =& N_{i,R} +N_{i,L} &  \xrightarrow{\,R,L\,} & RN_{i,R} +LN_{i,L}
\end{array}
\end{equation}
 Concerning parity one finds that $\tq q$ and $\tq \gamma_5 q$ transform as a scalar and a pseudo scalar, respectively. Thus, both $N_1$ and $N_2$ transform as a Dirac spinor under parity, $N_i \rightarrow \gamma_0 N_i$.
\begin{equation}\label{NTrafoP}
\begin{array}{rclcl}
N_{i} & =& N_{i,R} +N_{i,L} &  \xrightarrow{\, P \,} & \gamma_0(N_{i,L} +N_{i,R})\,.
\end{array}\end{equation}

So far we considered local interpolating fields only. In lattice QCD so-called smeared interpolators are very often used, mainly to suppress excited-state contributions in the correlation function. Smeared nucleon interpolating fields are formed as in \pref{DefN12} but with the local quark fields replaced by smeared ones, which are generically of the form\footnote{We use a continuum notation here. In lattice QCD the integral is replaced by a sum over the lattice points.} 
\begin{equation}
q_{\rm sm} (x) = \int {\rm d^4}y K(x - y) q(y)
\end{equation}
with some gauge covariant kernel $K(x-y)$ which is essentially zero for $|x-y|$ larger than some ``smearing radius'' $R$. The kernel depends on the details of the smearing procedure. Gaussian and exponential smearing \cite{Gusken:1989ad,Gusken:1989qx,Alexandrou:1990dq} is local in time and the kernel contains a delta function in the euclidean time coordinate. In contrast, the gradient flow \cite{Luscher:2013cpa} is a truly four-dimensional smearing. 

What matters here are the transformation properties of the smeared quark fields. Provided the kernel is diagonal in spinor space (as it is for Gaussian smearing and the gradient flow) the smeared quark fields transform just as the unsmeared ones under parity and global chiral transformations.  Consequently, also the nucleon interpolating fields formed with the smeared quark fields transform according to \pref{NTrafoLR} and \pref{NTrafoP}, just as their local counterparts. Since the symmetry properties of the interpolating fields essentially determine their expression in ChPT we can already conclude that both local and smeared interpolating fields are mapped onto the same effective operator, differing in their values for the LECs only. We come back to this issue in section \ref{sseceffFields}. 

The two interpolating fields in eq.\  \pref{DefN12} were originally discussed in the context of QCD sum rule calculations, and in lattice QCD simulations the discretized version of these continuum interpolaters are used.  An alternative approach for the construction of baryonic operators starts directly from the irreducible representations of the cubic group of the space-time lattice \cite{Basak:2005aq}. In order to discuss this type of operators one first has to perform a mapping to the Symanzik effective theory, the leading part being continuum QCD followed by corrections proportional to powers of the lattice spacing \cite{Symanzik:1983dc}. To the lattice operators whose leading Symanzik term is given by the interpolating fields in \pref{DefN12} the following discussion equally applies up to corrections proportional to the lattice spacing.\footnote{Lattice artifacts are of course also present in lattice simulations that use the discretized expressions of $N_1$ or $N_2$.}

Finally, non-relativistic interpolators can be constructed from the relativistic ones if the standard non-relativistic (Dirac-)representation for the $\gamma$-matrices is used \cite{Billoire:1984jm}. In appendix \ref{appendixA} we discuss briefly the correlation function of the non-relativistic limit of $N_1$, which is rather easily obtained from the result using the relativistic $N_1$.

\section{\label{secChPT} The nucleon two-point correlators in ChPT}
\subsection{\label{ssecLag} The chiral Lagrangian}
The framework for our calculations is covariant baryon chiral perturbation theory \cite{Gasser:1987rb,Becher:1999he}.\footnote{A thorough and pedagogical introduction to the subject can be found in Ref.\ \cite{Scherer:2012xha}, for example.} In this section we summarize a few relevant formulae since we work in euclidean space time and most references assume the Minkowski metric.

We consider the chiral effective Lagrangian\footnote{The superscripts denote the low-energy dimensions of these lagrangians, i.e.\ they count the number of derivatives  and the power of quark mass terms \cite{Gasser:1987rb}.}
\begin{equation}\label{effLag}
{\cal L}_{\rm eff}={\cal L}_{N\pi}^{(1)} + {\cal L}_{\pi\pi}^{(2)}\,.
\end{equation}
Here ${\cal L}_{\pi\pi}^{(2)}$ is the standard two-flavor mesonic chiral Lagrangian to leading order \cite{Gasser:1983ky,Gasser:1983yg}. According to the conventions used here it reads
\begin{equation}
{\cal L}_{\pi\pi}^{(2)} = \frac{f^2}{4} \Tr[\partial_{\mu}U\partial_{\mu}U^{\dagger}] +\frac{f^2B}{2}\Tr[{\cal M}(U+U^{\dagger})]\,.
\end{equation}
$f,B$ are the standard LO LECs related to the pion decay constant and chiral condensate in the chiral limit.\footnote{Our conventions correspond to $f_{\pi}= 92.2$~MeV.} The pion fields are contained in the field $U$ according to
\begin{equation}
U(x) = \exp\left[\frac{i}{f}\pi^a(x)\sigma^a\right],
\end{equation}
with the usual Pauli matrices $\sigma^a$. ${\cal M}$ denotes the quark mass matrix. With equal quark masses $m$ for the up and down quark it is proportional to the unit matrix. In that case all three pions have the same mass which to LO is related to the quark mass via $\mpi^2= 2Bm$. 

The second part in the chiral lagrangian \pref{effLag} contains the nucleon fields and their coupling to the pions,
\begin{equation}\label{LNpi}
{\cal L}_{N\pi}^{(1)}=\overline{\Psi} \Big(\slashed{D}+\mN -i \frac{g_A}{2}\slashed{u}\gamma_5\Big)\Psi\,.
\end{equation}
The  fields $\Psi=(p,n)^T$ and $\overline{\Psi}=(\overline{p},\overline{n})$ 
denote the nucleon fields with two Dirac spinors for the proton $p$ and the neutron $n$. 
$\mN$ and $g_A$ are the nucleon mass and the axial-vector coupling constant in the chiral limit. Since we assume isospin symmetry the proton and the neutron are mass degenerate. 

The pion fields enter ${\cal L}_{N\pi}^{(1)}$ via the field $u_{\mu}$, the so-called {\em chiral vielbein}, defined by
\begin{equation}
u_{\mu} = i[u^{\dagger}\partial_{\mu}u - u \partial_{\mu}u^{\dagger}],\quad u(x)\,=\,\sqrt{U(x)}\,.
\end{equation}
A second source of pion-nucleon coupling stems from the covariant derivative $\slashed{D}=\gamma_{\mu}D_{\mu}$ in \pref{LNpi}, with
$ D_{\mu}\Psi = \Big(\partial_{\mu} + \Gamma_{\mu}\Big)\Psi$ and 
 $\Gamma_{\mu}=\left[u^{\dagger}\partial_{\mu}u + u \partial_{\mu}u^{\dagger}\right]/2$. 

\begin{table}[tbdp]
\begin{center}
\begin{tabular}{c|cccccc}
\hline\hline
&\,&  $\Psi$&\,\,\, & $u$ & \,\,\,& $u_{\mu}$\\ 
\hline
\,$R\otimes L$ \, &&$K\Psi$  && $RuK^{\dagger}  = K u L^{\dagger} $ && $K u_{\mu} K^{\dagger} $\\
$P$  && $\gamma_0\Psi$ && $u^{\dagger}$ &&$ (- 1)^{\delta_{\mu 0}} u_{\mu} $\\
\hline
\end{tabular}
\end{center}
\caption{\label{tab:SymFields}Transformation behavior of the nucleon and pseudo scalar fields under chiral and parity transformations (see Ref.\ \cite{Wein:2011ix}). The SU(2) matrix $K$ appearing in the first row is defined by the transformation law of $u$ such that $u^2=U$ transforms in the standard way.}
\end{table}

The construction of the chiral lagrangian is based on the symmetry properties of the underlying QCD lagrangian, which ${\cal L}_{\rm eff}$ needs to reproduce. The transformation behavior of the nucleon and pseudo scalar fields under chiral and parity transformations is briefly summarized in table \ref{tab:SymFields} (for details see Ref.\ \cite{Scherer:2012xha}, for example).

Expanding $u_{\mu}$ and $\Gamma_{\mu}$  we obtain pion-nucleon interaction terms with various numbers of pion fields.  Since $u_{\mu}$ is parity-odd and $\Gamma_{\mu}$ is parity-even the leading interaction term with one pion field only stems from $u_{\mu}$ and reads
\begin{equation}\label{Lintexpand}
{\cal L}_{\rm int, LO}^{(1)} = \frac{ig_A}{2f}\overline{\Psi}\gamma_{\mu}\gamma_5\sigma^a \Psi \, \partial_{\mu} \pi^a\,.
\end{equation}
This interaction term couples two axial vectors to obtain a Lorentz scalar. In addition, isospin symmetry is preserved. 

The chiral Lagrangian incorporates a derivative expansion and the chiral dimension counts the number of derivatives  and powers of the quark mass. The complete list of terms through fourth order is given in \cite{Fettes:2000gb}. For the purpose of this paper, however,  the term in \pref{Lintexpand} is sufficient.

For the perturbative calculation in section \ref{ssecNpiCont} we need the propagators for the nucleon and the pion in position space.  The  pion propagator is the same as in Ref.\ \cite{Bar:2012ce},
\begin{equation}\label{scalprop}
G^{ab}(x,y)=   \delta^{ab}L^{-3}\sum_{\vec{p}} \frac{1}{2 E_{\pi}} e^{i\vec{p}(\vec{x}-\vec{y})} e^{-E_{\pi} |x_0 - y_0|}\,,
\end{equation} 
with 
pion energy $E_{\pi} =\sqrt{\vec{p}^2 +\mpi^2}$. The nucleon propagator $S^{ab}_{\alpha\beta}(x,y)$ is also easily derived from the quadratic term in \pref{LNpi}, and it reads
\begin{equation}\label{NucleonProp}
S_{\alpha\beta}^{ab}(x,y)=   \delta^{ab} L^{-3}\sum_{\vec{p}} \frac{Z_{p,\alpha\beta}^{\pm}}{2E_N} e^{i\vec{p}(\vec{x}-\vec{y})} e^{-E_N |x_0 - y_0|}\,.
\end{equation}
$a,b$ and $\alpha\beta$ refer to the isospin and Dirac indices, respectively. The factor $Z^{\pm}_{\vec{p}}$  (spinor indices suppressed) in the numerator is defined as
\begin{equation}
Z_{\vec{p}}^{\pm}=-i\vec{p}\cdot\vec{\gamma} \pm E_N \gamma_0+\mN\,, 
\end{equation}
where the $+$ ($-$) sign applies to $x_0 > y_0$ ($x_0 < y_0$), and the nucleon energy $E_N=\sqrt{\vec{p}^2 +\mN^2}$ involves the nucleon mass. 
The sum in both propagators runs over the discrete spatial momenta that are compatible with periodic boundary conditions, i.e.\ 
$\vp=2\p\vn/L$  with $\vn$ having integer-valued components.

\subsection{\label{sseceffFields} The chiral expansion of the interpolating fields}

The construction of the nucleon operators in baryon ChPT follows the standard procedure.  
Based on the symmetry properties of the operators on the quark level we write down the most general expression in the effective theory that has the same symmetry properties. This has essentially been done in Ref.\ \cite{Wein:2011ix} and we summarize the results needed in the following.\footnote{Instead of $N_{1,2}$ two other operators, related to $N_{1,2}$ by Fierz identities, are considered in Ref.\ \cite{Wein:2011ix}. However, the chiral expansion is essentially the same.} 

The nucleon operators in the effective theory needs to transform as given in \pref{NTrafoLR} and \pref{NTrafoP} under chiral and parity transformations. Basically, the nucleon operators are a sum of a right- and left-handed spinor and can be written as (we follow closely the notation introduced in Ref.\ \cite{Wein:2011ix})
\begin{equation}\label{Neff}
N= \sum_{n}\sum_k^{i_{n}} \alpha_k^{(n)} \left(N^{(n)}_{k,R} +  N^{(n)}_{k,L}\right)\,.
\end{equation}
$N^{(n)}_{k,R}$ and  $N^{(n)}_{k,L}$ are operators with low-energy dimension $n$. $i_n$ denotes the number of operators with chiral dimension $n$, which are labelled by the index $k$.  Under chiral and parity transformations the fields in \pref{Neff} transform according to 
\begin{equation}\label{NeffTrafo}
\begin{array}{rcl}
N^{(n)}_{k,R} +N^{(n)}_{k,L}    &  \xrightarrow{\,R,L \,}  & R\, N^{(n)}_{k,R} +L\, N^{(n)}_{k,L} ,\\[0.4ex]
N^{(n)}_{k,R} +N^{(n)}_{k,L}    &  \xrightarrow{\,\,\,P \,\,\,}  & \gamma_0 (N^{(n)}_{k,L} + N^{(n)}_{k,R})\, . 
\end{array}
\end{equation}
Each term on the right hand side of \pref{Neff} comes with its own LEC $\alpha_k^{(n)}$, and it is parity that relates the coefficients of the right- and left-handed contributions. 

An incomplete list of operators through chiral dimension two can be found in Ref.\ \cite{Wein:2011ix}. For convenience we reproduce the ones through $n=1$ in table \ref{table:Nop}. It is straightforward to check that these operators satisfy the transformation laws given in \pref{NeffTrafo}.

\begin{table}[tbdp]
\begin{center}
\begin{tabular}{ccccccc}
\hline\hline
$n$ &\,& $k$&\,\,\, & $N^{(n)}_{k,R}$ & \,\,\,& $N^{(n)}_{k,L}$\\ 
\hline
0 &&1 && $u P_+ \Psi$ && $u^{\dagger} P_- \Psi$\\
1 && 1 && $u u_{\mu} P_+ \gamma_{\mu} \Psi$ &&$ - u^{\dagger} u_{\mu} P_- \gamma_{\mu} \Psi $\\
1 && 2 && $u u_{\mu} P_+  D_{\mu} \Psi $&&$ - u^{\dagger} u_{\mu} P_- D_{\mu} \Psi$ \\
\hline
\end{tabular}
\end{center}
\caption{\label{table:Nop}Low-energy operators for the nucleon interpolating fields through chiral dimension one (see Ref.\ \cite{Wein:2011ix}).}
\end{table}

Note that $N$ in \pref{Neff} does not carry an index $i$ that would refer explicitly to one of the two interpolating fields defined in \pref{DefN12}. We dropped this index because the chiral expansion for both operators is the same due to their similar transformation behavior. The only difference are different values for the LECs in the chiral expansion. In order to keep our notation simple we suppress an additional index at the operator and the LECs in the following. 

Similarly, for the smeared interpolating fields discussed at the end of section \ref{ssect:QCDinterpolaters} we also find the same effective operator \pref{Neff} with different LECs. However, one qualification has to be made. Smeared interpolators with some ``size'' are mapped onto the pointlike nucleon field in the chiral effective theory. For this to be a good approximation the smearing radius needs to small compared to the Compton wave length of the pion. Provided this condition is met the pions do not distinguish between smeared and pointlike interpolating fields.\footnote{A concrete example is given in Ref.\ \cite{Bar:2013ora} where ChPT for some observables based on the gradient flow has been constructed.}

At lowest low-energy dimension only one operator contributes. Expanding $u,u^{\dagger}$ in powers of pion fields and keeping only the terms up to linear order we obtain for $N$ the expression
\begin{equation}\label{Neffexp}
N(x)=  \tilde{\alpha} \left(\Psi(x) + \frac{i}{2f} \pi(x) \gamma_5\Psi(x)\right)\,,\qquad \tilde{\alpha}\,=\,4\alpha_0^{(0)}\,.
\end{equation}
The first LO term is proportional to the nucleon field $\Psi$, as expected. The second NLO term (suppressed by $1/f$) involves a nucleon-pion coupling that will contribute to the two-particle nucleon-pion terms \pref{tpcontr} in the nucleon correlation function. 

\subsection{\label{ssecNpiCont} Perturbative expansion of the correlation functions}

\begin{figure}[tp]
\begin{center}
\includegraphics[scale=0.4]{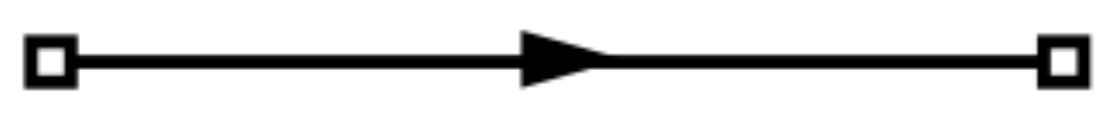}\\
a)\\[3ex]
\includegraphics[scale=0.4]{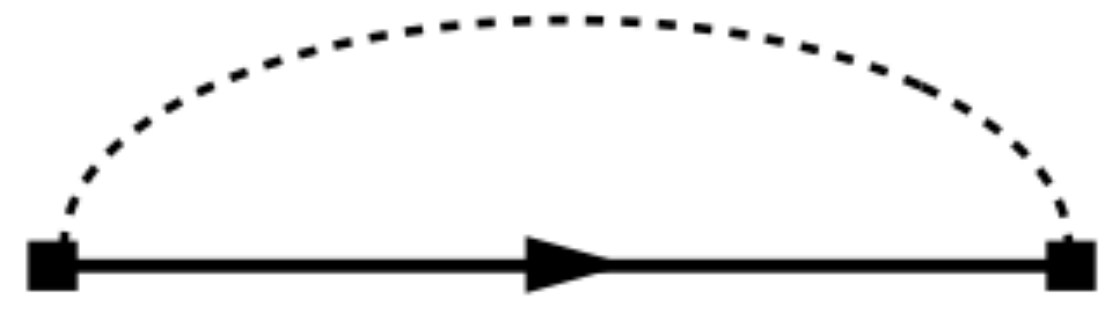}\\
b)\\[3ex]
\includegraphics[scale=0.4]{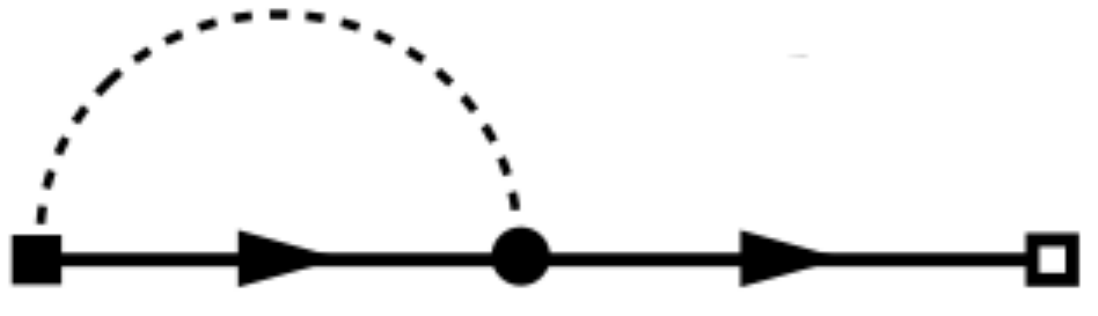}\hspace{1cm}\includegraphics[scale=0.4]{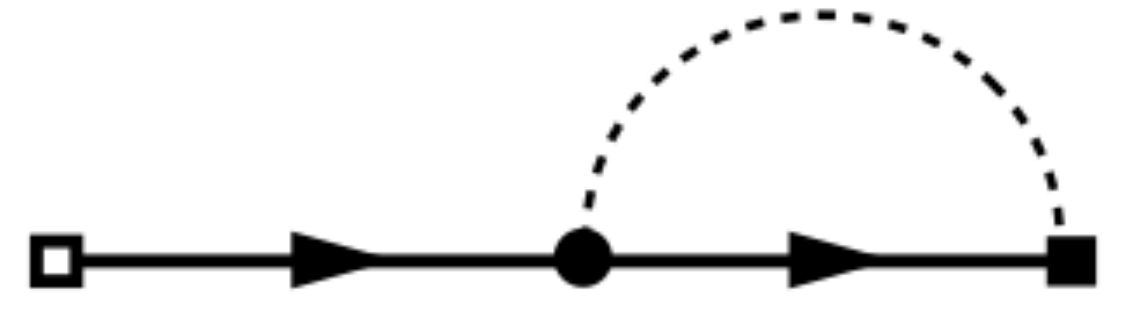}\\
c)\hspace{5cm} d)\\[3ex]
\includegraphics[scale=0.4]{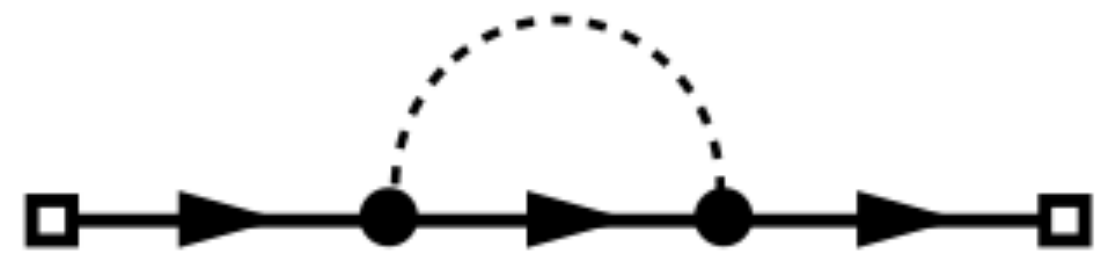}\\
e)
\caption{Feynman diagrams for the nucleon correlation function. The squares represent the nucleon operator at times $t$ and $0$, where the open and solid squares denote the leading and next-to-leading order terms given in \pref{Neffexp}. The circles represent a vertex insertion at an intermediate space time point; and an integration over this point is implicitly assumed. The solid and dashed lines represent nucleon and pion propagators, respectively. }
\label{fig:diagrams}
\end{center}
\end{figure}

We are now in the position to compute the correlation functions \pref{DefNCorr} perturbatively within the chiral effective theory. The leading contribution is obtained by taking into account the LO term in \pref{Neffexp} for $N$ and $\overline{N}=N^{\dagger} \gamma_0$.   Since these fields are proportional to the nucleon fields $\Psi,\overline{\Psi}$ the LO contribution is essentially the nucleon propagator. In terms of Feynman rules in position space this contribution is represented by the Feynman diagram in figure \ref{fig:diagrams}a. Taking into account \pref{NucleonProp} for the nucleon propagator the LO results for the correlators are easily obtained, 
\begin{eqnarray}\label{CorrLO}
C_{+,N}(t)&=& 2|\tilde{\alpha}|^2 e^{-M_N t}\,,\qquad C_{-,N}(t)\,=\, 0\,.
\end{eqnarray}
These results are a single-particle state contributions to the correlation function, and by comparing with \pref{spcontr} we can read off the LO relation between the vacuum-to-nucleon matrix element and the LEC $\tilde{\alpha}$,
\begin{equation}
|\langle 0|N_{+}(0)|N_{+}(\vp=0)\rangle|^2 = 4M_N |\tilde{\alpha}|^2\,.
\end{equation}
$C_{-,N}$ in \pref{CorrLO} vanishes at this order because our effective theory does not contain the negative parity nucleon as a degree of freedom. 

The diagrams in Figs.\ \ref{fig:diagrams}b - \ref{fig:diagrams}e form the leading contribution to the two-particle nucleon-pion part of the correlation function. The dashed line represents the pion propagator, which, together with the nucleon propagator, leads to terms with the expected exponential fall-off with $E_{\rm tot} = E_N+E_{\pi}$.\footnote{Fig.\ \ref{fig:diagrams}e also contains a contribution with  a time dependence proportional to $t\exp(-M_N t)$. This results in the renormalization of the nucleon mass and can be ignored for our purposes.} 

The calculation of the diagrams \ref{fig:diagrams}b - \ref{fig:diagrams}e is straightforward, and the final results can be compactly written as 
\begin{eqnarray}
C_{+,{N\pi}}(t)& = &2|\tilde{\alpha}|^2\frac{3}{8(fL)^2 mL} \sum_{\vec{p}} 
\frac{m}{E_{\pi}} \frac{E_N-M_N}{2 E_N}\left[1 -g_A \frac{E_{\rm tot}+M_N}{E_{\rm tot}-M_N}\right]^2
 e^{-E_{\rm tot} |t|}\,,\label{Cnpplus}\\
C_{-,{N\pi}}(t)& = &2|\tilde{\alpha}|^2\frac{3}{8(fL)^2 mL} \sum_{\vec{p}} 
\frac{m}{E_{\pi}}\frac{E_N+M_N}{2E_N}\left[1 -g_A \frac{E_{\rm tot}-M_N}{E_{\rm tot}+M_N}\right]^2
 e^{-E_{\rm tot}|t| }\,.\label{Cnpminus}
\end{eqnarray}
Note that the $\vec{p}=0$ term  in $C_{+,{N\pi}}(t)$ correctly vanishes, as required by parity.

The overall factor $2|\tilde{\alpha}|^2$ has its origin in the appearance of $|\tilde{\alpha}|^2$ as an overall factor for the two terms in \pref{Neffexp}. This implies that the relative size of the two-particle state contributions in \pref{Cnpplus}, \pref{Cnpminus} and the one-particle state contribution in \pref{CorrLO} does not contain the LEC $\tilde{\alpha}$ associated with the nucleon interpolating field. It only depends on the LECs $f$ and $g_A$ of the effective action. However, this is true to LO only. Taking into account the higher order operators in table \ref{table:Nop} their LECs will not cancel in the ratio. 

Recall that the chiral expansion for the nucleon operators $N_1$ and $N_2$ are the same, the only difference being different LECs. Restoring the label $i=1,2$ in the LEC $\tilde{\alpha}$ the results given above (with $|\tilde{\alpha}|^2$ replaced by $|\tilde{\alpha}_i|^2$) refer to the correlation functions where the same operator is used for both source and sink. In case of the correlator with $N_1(x)$ and $\overline{N}_2(0)$ we simply need to replace $|\tilde{\alpha}|^2$ by $\tilde{\alpha}_1{\tilde{\alpha}_2^{*}}$. Since the same combination appears as an overall factor in the single-particle and two-particle state contributions it still drops out in the ratio of these two contributions. The same statement applies to smeared operators.

The summation in \pref{Cnpplus}, \pref{Cnpminus} is  over all lattice momenta compatible with periodic boundary conditions. Those that are related by the symmetries of the spatial lattice lead to the same contribution, hence it is convenient to  sum over the absolute value $p=|\vec{p}|$. Imposing periodic boundary conditions the absolute value can assume the values $p_n=(2\pi/L)\sqrt{n}$, $n\equiv n_1^2+n_2^2+n_3^2$, with the $n_k$ being integers, and the sums in the results given above are replaced according to
\begin{equation}\label{redsum}
\sum_{\vec{p}} \longrightarrow \sum_{p_n} m_n\,.
\end{equation}
The multiplicities $m_n$ count the number of vectors $\vec{p}$ with the same $p_n$. Multiplicities for $n\leq 20$ are given in Ref.\ \cite{Colangelo:2003hf} (for convenience we summarize the first eight in table \ref{tabledn}).

\begin{table}[tbdp]
\begin{center}
\begin{tabular}{lrrrrrrrrr}
\hline\hline
$n$ & 0 & 1 & 2 & 3 & 4 & 5 & 6 & 7 & 8\\ 
$m_n$ & \phantom{1}1 & \phantom{1}6 & 12 & \phantom{1}8 & \phantom{1}6 & 24 & 24 & \phantom{0}0 & 12\\
\hline
\end{tabular}
\end{center}
\caption{\label{tabledn}Multiplicities $m_n$ in eq.\ \pref{redsum} for $n\le 8$ (see Ref.\ \cite{Colangelo:2003hf}).}
\end{table}

\subsection{\label{ssecResults} Final Results}

Adding the two results in \pref{CorrLO} and \pref{Cnpplus} the positive parity correlation function can be written as
\begin{equation}\label{Cplusfinal}
C_{+}(t)= 2|\tilde{\alpha}|^2 e^{-M_N t}\left[ 1+ \sum_{p_n} c^{+}_n e^{-(E_{{\rm tot},n} -\mN) t}\right]\,,
\end{equation}
where we introduced new dimensionless coefficients 
\begin{eqnarray}
c^{+}_n &=& \frac{3m_n}{8(fL)^2 E_{{\pi,n}}L} h^{+}_n\,,\label{cnp}\\
  h^{+}_n &=&  \frac{E_{N,n}-M_N}{2 E_{N,n}}\left[1 -g_A \frac{E_{{\rm tot},n}+M_N}{E_{{\rm tot},n}-M_N}\right]^2\,.\label{hnp}
\end{eqnarray}
Result \pref{Cplusfinal} corresponds to the example we discussed briefly in the introduction. The coefficients $c_n^+$ are equal to $b_n/b_0$ and all nucleon-pion state contributions are given. 

We already mentioned that the positive parity two-point correlator was independently calculated in Ref.\ \cite{Tiburzi:2015tta} using HBChPT \cite{Georgi:1990um,Jenkins:1990jv}. In that non-relativistic formulation one drops the antinucleon degrees of freedom and the dispersion relation of the heavy nucleon is non-relativistic. If we expand $E_{N,n}\approx M_N+ p_n^2/2M_N$ in \pref{hnp} and drop all but the dominant terms we find $h_n^+ \approx g_A^2 p_n^2/E_{\pi}^2$ and reproduce the result in \cite{Tiburzi:2015tta}.  

The negative parity channel is slightly different since there is no  single particle state contribution stemming from the nucleon. The leading single-particle state contribution comes from the negative parity partner $N^*$ which is not a degree of freedom in our effective theory. And even if we included it explicitly the coupling of the interpolating field $N^-$ to the $N^*$ would come with a LEC unrelated to the LEC $\tilde{\alpha}$ that enters result \pref{Cnpminus}. 

The dominant contribution in our result for $C_-$ stems from the nucleon-pion state with the nucleon and the pion at rest. Taking this contribution out of the sum we arrive at the form
\begin{equation}\label{Cminusfinal}
C_{-}(t)= 2|\tilde{\alpha}|^2 \frac{3}{8(fL)^2 \mpi L} e^{-(\mN + \mpi) t}\left[ 1+ \sum_{p_{n}\neq 0} c^{-}_n \,e^{-(E_{{\rm tot},n} -\mN - \mpi) t}\right]\,,
\end{equation}
with a coefficient 
\begin{eqnarray}
c^{-}_n &=& m_n\frac{\mpi}{E_{\pi,n}} h^{-}_n\,,\label{cnm}\\
h^{-}_n &=&  \frac{E_{N,n}+M_N}{2 E_{N,n}}\left[1 -g_A \frac{E_{{\rm tot},n}-M_N}{E_{{\rm tot},n}+M_N}\right]^2\,.
\end{eqnarray}

Note that the results for the two parity channels are proportional to the unknown LEC $|\tilde{\alpha}|^2$. Thus, taking the ratio $C_{-}(t)/C_{+}(t)$ this constant drops out and the only LECs contributing to this ratio are $f$ and $g_A$ of the LO chiral lagrangian.

\subsection{\label{ssecNumEst} Numerical estimates}

Before trying to estimate the impact of the nucleon-pion-state contribution to the two-point function it is necessary to discuss the conditions for the applicability of the results derived in the last section. 

ChPT is an expansion in the pion mass and momentum. Both need to be small compared to the chiral symmetry breaking scale $\Lambda_{\chi}$, which is typically identified with $4\pi f_{\pi}$. In a finite spatial volume with periodic boundary conditions the pion momenta are discrete and the condition for the applicability of the chiral expansion reads \cite{Colangelo:2003hf}
\begin{equation}\label{boundpn}
\frac{p_n}{\Lambda_{\chi}} = \frac{M_{\pi}}{2f_{\pi}} \frac{\sqrt{n}}{M_{\pi} L}  \ll 1\,.
\end{equation}
Even though the pion mass cancels on the right hand side we prefer this form since lattice QCD configurations are often characterized in terms of the pion mass and $M_{\pi}L$. Given these two numbers and the pion decay constant $f_{\pi}\approx 90$MeV eq.\ \pref{boundpn} provides a bound on the pion momenta and the label $n$. 
Table \ref{tab:expansion} lists a few representative values that approximately match the parameters in present-day lattice simulations.\footnote{For the calculation of the light hadron spectrum in \cite{Durr:2008zz} the BMW collaboration generated a lattice ensemble with $M_{\pi}\approx 190$MeV with $M_{\pi}L\approx 3.9$. The PACS collaboration has recently reported results obtained on a $96^4$ lattice with $M_{\pi}\approx 147$MeV and $M_{\pi}L\approx 6$ \cite{UkitaLat2015}. Finally, the FERMILAB and MILC collaborations \cite{Bazavov:2014wgs} have generated a lattice ensemble with $M_{\pi}\approx 128$MeV and $M_{\pi}L\approx 3.9$.}
\begin{table}[tbd]
\begin{center}
\begin{tabular}{cc|cc|c}
$M_{\pi}$ & $\,M_{\pi}L\,$ & $\,M_{\pi}/\Lambda_{\chi}\,$  & $\,p_n/\Lambda_{\chi}\,$ & $\,n_{\rm max}$ \\[0.8ex]
\hline
200 & 4 & $\simeq 1/6$ & $\simeq \sqrt{n}/4 $ & 1\\
150 & 4 & $\simeq 1/8$ & $\simeq \sqrt{n}/5 $ & 2 \\
130 & 4 & $\simeq 1/9$ & $\simeq \sqrt{n}/6 $ & 3 \\
150 & 6 & $\simeq 1/8$ & $\simeq \sqrt{n}/8 $ & 5 \\
\end{tabular}
\end{center}
\caption{\label{tab:expansion}  The chiral expansion parameters $M_{\pi}/\Lambda_{\chi}$ and $p_n/\Lambda_{\chi}$ for various pion masses and spatial extensions $L$. $n_{\rm max}$ in the last column stems from the condition $p_{n_{\rm max}}/\Lambda_{\chi}\approx0.3$}
\end{table}
The expansion parameter $M_{\pi}/\Lambda_{\chi}$ is sufficiently small for pion masses smaller than $200$MeV. The situation is less favourable for $p_n/\Lambda_{\chi}$. In order to have at least theoretically a  chiral expansion at all the expansion parameter $p_n/\Lambda_{\chi}$ should be reasonably smaller than 1. If we restrict ourselves to momenta satisfying $p_n/\Lambda_{\chi} \le 0.3$  we find values for $n_{\rm max}$ ranging between 2 and 5. For the coefficients $c_n^+$ with $n\lesssim n_{\rm max}$ we expect the chiral expansion to be applicable.
 
Besides the question of applicability there is the question of how rapidly the chiral expansion converges. The smaller $n$ and $p_n$ is the better the chiral expansion behaves.  For $p_n/\Lambda_{\chi} \simeq0.3$ one does not expect a fast rate of convergence, and the LO results for the coefficients $c_n^{+}$ associated with these rather high momenta will probably receive rather large higher order corrections. Definite statements about these higher order corrections are difficult to make without having done the calculation, but as a rough error estimate we may allow for a 50\% error. The error will be smaller for the coefficients associated with the smaller momenta. Note that the exponential suppression due to the exponential $\exp[-(E_{{\rm tot},n} -\mN) t]$  is stronger for the contributions with larger momentum,  so the contributions with a larger uncertainty in the two-point function are more suppressed.   

Another reason to constrain the momenta $p_n$ and $n$ stems from the requirement that the nucleon-pion-state energy should be sufficiently well separated form the first one-particle resonance energy. If that is not the case one expects large mixing between these two states that significantly alter the coefficents $c_n^{+}$ \cite{Luscher:1991cf}.  In practice, however, this seems to be almost automatically satisfied once the momenta are constrained by the bound \pref{boundpn}, because the energy of the first resonance is about 0.5GeV higher than the nucleon mass. For example, for the values $n_{\rm max}$  given in table \ref{tab:expansion} the total energy of the nucleon-pion state is at least 100MeV smaller than the expected resonance energy, and the energy gap for the states with $n<n_{\rm max}$ is even smaller. We therefore expect mixing effects to be negligible, at least within the uncertainties the LO results presented here are afflicted with.

In lattice simulations one usually computes the effective nucleon mass, defined as the negative time derivative of $\log C_+(t)$.  With \pref{Cplusfinal} we obtain
\begin{eqnarray}
M_{N,{\rm eff}} & =&  M_N \left[ 1+ \sum_{p_n} d^{+}_n e^{-(E_{{\rm tot},n} -\mN) t}\right]\,,\label{Meff}\\ 
d_n^+ & =&  c_n^+\left[\frac{E_{{\rm tot},n}}{M_N}-1\right]\,.
\end{eqnarray}
\begin{figure}[t]
\begin{center}
\includegraphics[scale=0.45]{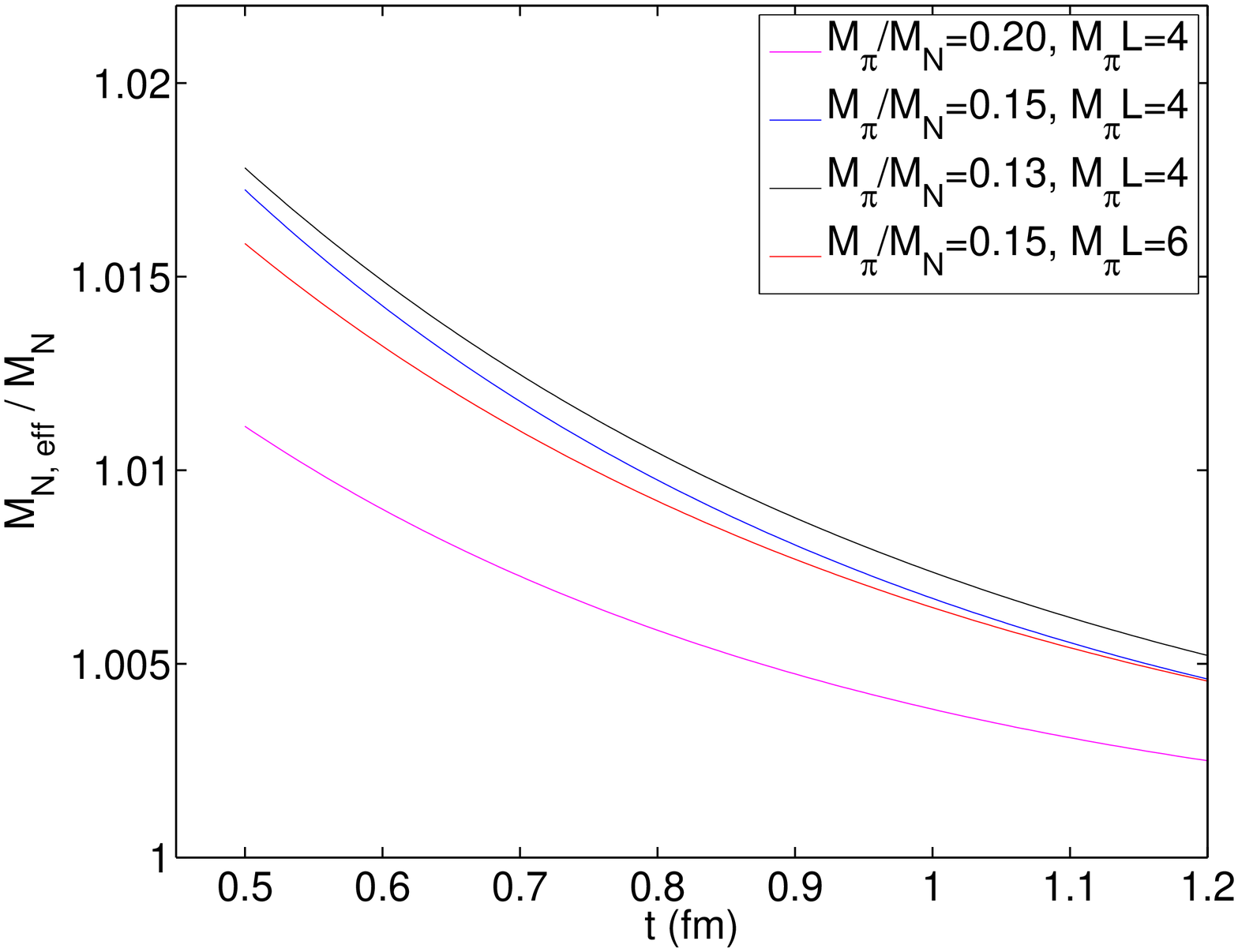}\\
\caption{$M_{N,{\rm eff}}/M_N$ as a function of the sink-source separation $t$ for the parameter sets in table \ref{tab:expansion}. }
\label{fig:contMeff}
\end{center}
\end{figure}
Figure \ref{fig:contMeff} shows the ratio $M_{N,{\rm eff}}/M_N$ as a function of $t$ for the four parameter sets in table \ref{tab:expansion}. The deviation from the constant value 1 is caused by nucleon-pion-state contribution. The sum over $p_n$ is truncated at $n_{\rm max}$ given in table \ref{tab:expansion}. 
We conclude that the nucleon-pion-state contribution to the effective mass is small, less than 2 percent for $t\ge0.5$fm, and dropping below 1 percent at $t\approx 1$fm. Note that the nucleon-pion-state contribution to the effective mass is significantly smaller than to the two-point function itself since the coefficients $d_n^+$ are smaller than the coefficients $c_n^+$. For the parameter sets of table \ref{tab:expansion} the factor $E_{{\rm tot},n}/M_N-1$ causing this suppression is less than about 0.4. 

Figure \ref{fig:contMeff} shows the results to leading order in the chiral expansion. As argued before we may allow for a 50\% error for the coefficents, which approximately results in a 50\% error in figure \ref{fig:contMeff}. Even with this uncertainty we can conclude that the nucleon-pion-state contribution to the effective mass is at the few-percent level.  Whether this plays a role in practice depends on the statistical errors in the lattice data and the source-sink separations accessible in the simulation. For statistical errors of  1\% and below the nucleon-pion-state contribution should be noticeable in the data, at least for euclidean times below $0.7$ fm. Obviously, actual lattice data needs to be analyzed with the results presented here before definite conclusions can be drawn.\footnote{We mention that Ref.\ \cite{Mahbub:2013bba} reports on the presence of a non-negligible nucleon-pion-state contribution to $C_-$. Unfortunately, a direct check of our results is not possible.}
 
\section{\label{secConcl} Concluding remarks}

The results we find for the nucleon-pion-state contribution to the nucleon correlation function is  small. 
Whether it is too small to be seen in lattice QCD data is not easy to answer and depends on various parameters of the simulation, among others on the range for the euclidean time and the size of the statistical errors in the data. Future physical-point simulations aiming at a one-percent error for the nucleon mass will probably be sensitive to the corrections studied here.

We repeatedly mentioned that the calculations in this paper are LO calculations. In principle, the higher order corrections can be calculated straightforwardly. Whether this is useful in practice is somewhat doubtful. The NLO contributions will depend on the LECs associated with the interpolating nucleon fields. Since their values are a priori unknown the NLO results are in a way less predictive than the LO results derived here. On the other hand, the LO results are universal for all the various nucleon interpolating fields considererd here. In order to discuss and describe differences caused by using different interpolating fields one has to go at least one order higher in the chiral expansion.

Here we studied the dominant nucleon-pion-state  contributions to the nucleon two-point function. The same framework can be used 
to  study the nucleon-pion-state contribution to higher $n$-point functions.  The nucleon axial charge, for example,  is obtained from the three-point function involving the axial vector current. In that case the contamination with excited and multi-particle states is expected to be larger, because more than one euclidean time difference are present in these correlation functions, and these are usually smaller than the source-sink separation in the two-point function. Preliminary results concerning the nucleon-pion-state  contributions in the determination of $g_{\rm A}$ can be found in \cite{Bar:2015zha}. Other interesting observables to study are the electromagnetic form factors and the quark momentum fraction in the nucleon, for instance. 

\vspace{2ex}
\noindent {\bf Acknowledgments}
\vspace{2ex}

Discussions with Tomasz Korzec and Stefan Schaefer are gratefully acknowledged. I thank Maarten Golterman and Rainer Sommer for their feedback on a first draft of this manuscript. 
\vspace{3ex}

\begin{appendix}

\section{Non-relativistic interpolating fields}
\label{appendixA}

As mentioned at the end of section \ref{ssect:QCDinterpolaters} non-relativistic interpolating nucleon fields can be constructed from their relativistic counterparts. Here we briefly discuss the minor modification this choice implies in case of $N_1$. It is sufficient to discuss the local interpolating field only, the generalization to the smeared fields is completely analogous to the discussion at the end of section \ref{ssect:QCDinterpolaters}.

In our discussion of the relativistic interpolating fields there was no need to specify a particular representation for the $\gamma$-matrices. For the non-relativistic limit it is useful to assume the Dirac representation with $\gamma_0$ being diagonal. The reason is that in this basis the projectors $\Gamma_{\pm} = (\gamma_0 \pm 1)/2$ can be used to decompose a quark spinor $q$ into its  {\em upper} (large) spinor component $q_u$ and its {\em lower} (small) spinor components $q_l$.  The latter vanish in the non-relativistic limit.

Decomposing $N_1$ into upper and lower components one finds that one contribution involves upper components only \cite{Leinweber:2004it}. Let us call this part $N_{1,{\rm nr}} = (\tilde{q}_{u} q_u)q_u$. This field completely breaks charge conjugation but it still transforms according to (2.8) and (2.9) under chiral transformations and parity. So this field is mapped onto the expression \pref{Neffexp} in the chiral effective theory but the nucleon interpolating field $\Psi$ in the terms of table \ref{table:Nop} are replaced by $\Gamma_+\Psi$.  Using this  we find the expression 
\begin{equation}\label{Nnreffexp}
N_{1,{\rm nr}}(x)=  \tilde{\alpha}_{\rm nr} \left(\Gamma_+\Psi(x) -\Gamma_{\!-} \frac{i}{2f} \pi(x) \gamma_5\Psi(x)\right)
\end{equation}
for the interpolating field in the chiral effective theory. In addition to a different LEC the parity projectors appear explicitly in this expression, in contrast to the result in \pref{Neffexp}. 

Suppose we compute the positive parity correlation function with this interpolating field. Due to the parity projector $\Gamma_{+}$ in the definition of the correlator only the first term in \pref{Nnreffexp} contributes, the second vanishes identically. In terms of Feynman diagrams this means that only diagram e) in fig.\ \ref{fig:diagrams} contributes. Dropping the contributions from the other diagrams in the result \pref{Cnpplus} is very simple, it amounts in dropping the 1 in $[1- g_A \ldots]^2$ on the right hand side of \pref{Cnpplus}. This is esily understood since diagram e) is the only one with two vertex insertions and therefore the only one proportional to $g_A^2$. This is the only modification there is, thus the final result for the correlator is the same as in \pref{Cplusfinal} provided the aforementioned modification is made in the coefficient $h_n^+$.  Note that dropping the 1 in $[1- g_A \ldots]^2$ makes the coefficient $h_n^+$ larger, even though the increase is not huge, at least for small energies where the contribution proportional to $g_A$ is much larger than 1.

A similarly simple modification is found in the negative parity correlator. In that case only the second term in \pref{Nnreffexp} provides a non-vanishing contribution to the correlator, thus only diagram b) contributes. This is accounted for by keeping only the 1 in $[1- g_A\ldots ]^2$ in \pref{Cnpminus} and \pref{Cminusfinal}, respectively.

\end{appendix}

\end{document}